\documentclass[article,twocolumn,onecolappendix,usenatbib,useAMS]{aastex61}

\usepackage{amssymb}
\usepackage{amsmath}
\usepackage{tabularx}
\usepackage{color}
\usepackage{graphicx}
\usepackage{rotating}
\usepackage{subfigure}

\hypersetup{linkcolor=red,citecolor=blue,filecolor=cyan,urlcolor=magenta}

\newcommand{\msun}{\ensuremath{M_\odot}}

\accepted{ApJ, 24 May 2018 }

\begin{document}

\def\apj{{ApJ}}                 
\def\apjl{\ref@jnl{ApJ}}                
\def\mnras{{MNRAS}}             
\def\apjl{{ApJ}}                
\def\apjs{l{ApJS}}               
\def\apss{{Ap\&SS}}             
\def\pasp{{PASP}}               
\def\aap{{A\&A}}                
\def\aapr{{A\&A~Rev.}}          

\title{The Stellar Halo of the Spiral Galaxy NGC 1560}
\correspondingauthor{Laura Greggio}
\email{laura.greggio@oapd.inaf.it}


\author{Laura Greggio}
\affiliation{INAF, Osservatorio Astronomico di Padova, Vicolo dell'Osservatorio 5 I-35122 Padova - ITALY}

\author{Renato Falomo}
\affiliation{INAF, Osservatorio Astronomico di Padova, Vicolo dell'Osservatorio 5 I-35122 Padova - ITALY}

\author{Riccardo Scarpa}
\affiliation{Instituto de Astrofisica de Canarias, C/O Via Lactea, s/n E38205 - La Laguna (Tenerife) - SPAIN}
\affiliation{Universidad de La Laguna, Dpto. Astrofsica, s/n E-38206 La Laguna (Tenerife) - SPAIN}

\begin{abstract}
We report on the detection of a stellar halo around NGC 1560, a $10^9 \msun$ spiral galaxy member of the Maffei group. We obtained deep images in the $r$ and $i$ bands using the 10.4m Gran Telescopio Canarias in a field centered at $\sim$ 3.7 arcmin (projected distance of 3.5 kpc) from the center of this galaxy. The luminosity function and the CMD show a clear excess of  stars with respect to the expected foreground level at magnitudes fainter than the RGB tip at the distance of NGC 1560. The color of the halo stars implies a metallicity of $Z \sim Z_\odot/50$, while their counts correspond to a stellar mass of $\sim 10^{7} \msun$ in the sampled region. Assuming a power law profile for the surface mass density of the halo, our data suggest a total stellar mass of $10^8 \msun$ for the halo of NGC 1560.
\end{abstract}

\keywords{}

\section{Introduction} \label{sec:intro}

According to models of galaxy formation in a hierarchical universe, the accretion and disruption of small mass satellite galaxies which get captured in the potential well of the main host result in the build up of extended stellar haloes around galaxies \citep[e.g.][]{Cooper+10}. This 
scenario is strongly supported by the detection of structures in galaxy haloes \citep[e.g.][] {Ferguson+02}, including signatures of  ongoing accretion taking place \citep{Foster+14, Martinez+15}. 
Structures in galaxy halos are usually detected in integrated light \citep{Martinez+10, Miskolczi+11, Merritt+16}, but only if their surface brightness is high enough. However, the accreted units are eventually completely dissolved, their stars making up a smooth, diffuse, low surface brightness stellar halo. For such component, star counts are more effective to detect and measure the properties (mass, age and metallicity) of the stellar halos, which  show up as an excess of sources above the general background. Adopting this tecnique several studies have mapped the smooth stellar halos around galaxies \citep{Mouhcine+05, Barker+09, Mouhcine+10,Tanaka+11, Greggio+14, Harmsen+17}. 
These studies mostly pertain to galaxies with mass comparable to the Milky Way, or more massive, a range in which the potential well of the host galaxy is deep, so that the accretion of several units is well justified.

 At the other mass end, extended stellar halos have been revealed also in many dwarf galaxies \citep[see, e.g.][]{ Minniti+96, Stinson+09, McConnachie16}. In some cases the merging of two dwarf galaxies takes place, as clearly documented for NGC 4449 \citep{Martinez+12, Rich+12}, so that the low mass unit will eventually provide a stellar halo to the more massive accretor \citep{Bekki08}. However, this occurrence could be rare in the standard picture of hierarchical structure formation, and other mechanisms have been proposed to be responsible for the extended halo of dwarfs. In particular, hydrodynamical simulations of galaxy formation show that extended haloes might form around dwarf isolated galaxies due to the contraction of the star forming envelope, and to the ejection of stars formed in supernova driven shocks. The latter process is particularly  effective in the lower mass models, with the shallower potential well providing a weaker barrier to the supernovae driven outflows \citep{Stinson+09}.

In this scenario, low mass spiral galaxies  could be at the crossroad of these two paths of stellar haloes formation. Massive enough to inhibit a pronounced dispersion of their stars due to the supenovae driven outflows, but not massive enough to efficiently accrete smaller units. These galaxies could host the least conspicuous stellar halos, either in mass fraction, or extension, or both.  Thus, we can derive important clues on the general picture concerning the formation of stellar haloes investigating this class of galaxies to answer the questions: do low mass spirals host stellar haloes? What is their mass, extension, density profile?   
Do they resemble a scaled down version of the haloes of more massive spirals?

The most effective way to measure the properties of a diffuse, low surface brightness, stellar halo is via individual star counts. In stellar populations older than $\sim$ 2 Gyr the evolved stars gather on the RGB, forming a well recognizable feature on the CMD. The number of stars per unit mass of the parent stellar population on the upper magnitude of the RGB is almost independent of age from $\sim 2$ to $\sim 10$ Gyr \citep[e.g.][] {Book11}, making these stars effective and robust probes of the total underlying stellar mass. 

In this paper we present $i$ and $r$ band photometry of stars in a field located in the halo of NGC 1560 (Section 2), based on observations at the Gran Telecopio CANARIAS. The program aimed primarily at assessing whether NGC 1560 hosts a stellar halo, and in case at estimating its size.
In Section 3 we describe the observations and the data reduction procedure; in Section 4 we present the CMD and luminosity function of the stars in the NGC 1560 halo field; in Section 5 we summarize our results and discuss them in comparison to the properties of other galaxies, and to the predictions of galaxy formation models.

\section{The low mass spiral galaxy NGC 1560}

NGC 1560 is a late type spiral galaxy (Scd, Sdm) member of  the Maffei Group. The distance (3.27Mpc) to NGC 1560 was evaluated by \cite{Jacobs+09} from the location of the Tip of the Red Giant Branch (RGB)  on the Colour Magnitude Diagram (CMD) derived from HST photometry in $V$ and $I$. The object is located at low galactic latitude ($b$=16 deg), so that the foreground extinction is quite large ($E(B-V) \simeq$ 0.18).  This galaxy is relatively isolated: the two nearest companions are Cam A and Cam B at a distance of  $\sim$ 500 and $\sim$ 400 kpc, respectively \citep{Karachen+04}.

The stellar disk is viewed edge-on (inclination of $\sim 80 ^{\circ}$), and it is embedded in an extended HI disk, characterized by regular contours. From the analysis of the HI rotation curve \cite{Gentile+10} derived a total (dark + luminous) mass of $\sim 4 \times 10^{10}$ \msun, while they estimated a stellar mass of $\sim 5 \times 10^8$ \msun\ from I band photometry. 
 Therefore NGC 1560 is a very good candidate to investigate the properties of stellar haloes at the low mass end of spiral galaxies, also because of the favourable edge-on orientation which allows an unambigous sampling of the halo component. 
 
 On the Padova isochrones \footnote{stev.oapd.inaf.it/cgi-bin/cmd} \citep{Marigo+08} the RGB Tip of a 10 Gyr old stellar population is found at M$_i \simeq -3.25$ and $-3.47$, respectively for metallicities $Z$ = 0.004 and 0.0004. The foreground galactic extinction is of 0.32 and 0.43 mag in the $i$ and $r$ bands respectively. Correspondingly, we expect the RGB Tip of an old and metal poor stellar population in the halo of NGC 1560 to be found at $i \sim 24.5$.  In the following we will adopt this reference value to flag the presence of  RGB stars members of NGC 1560.

\section{Observations, data reduction and analysis} \label{sec:data}
\begin{figure*}
   \centering
  \includegraphics[width=13cm]{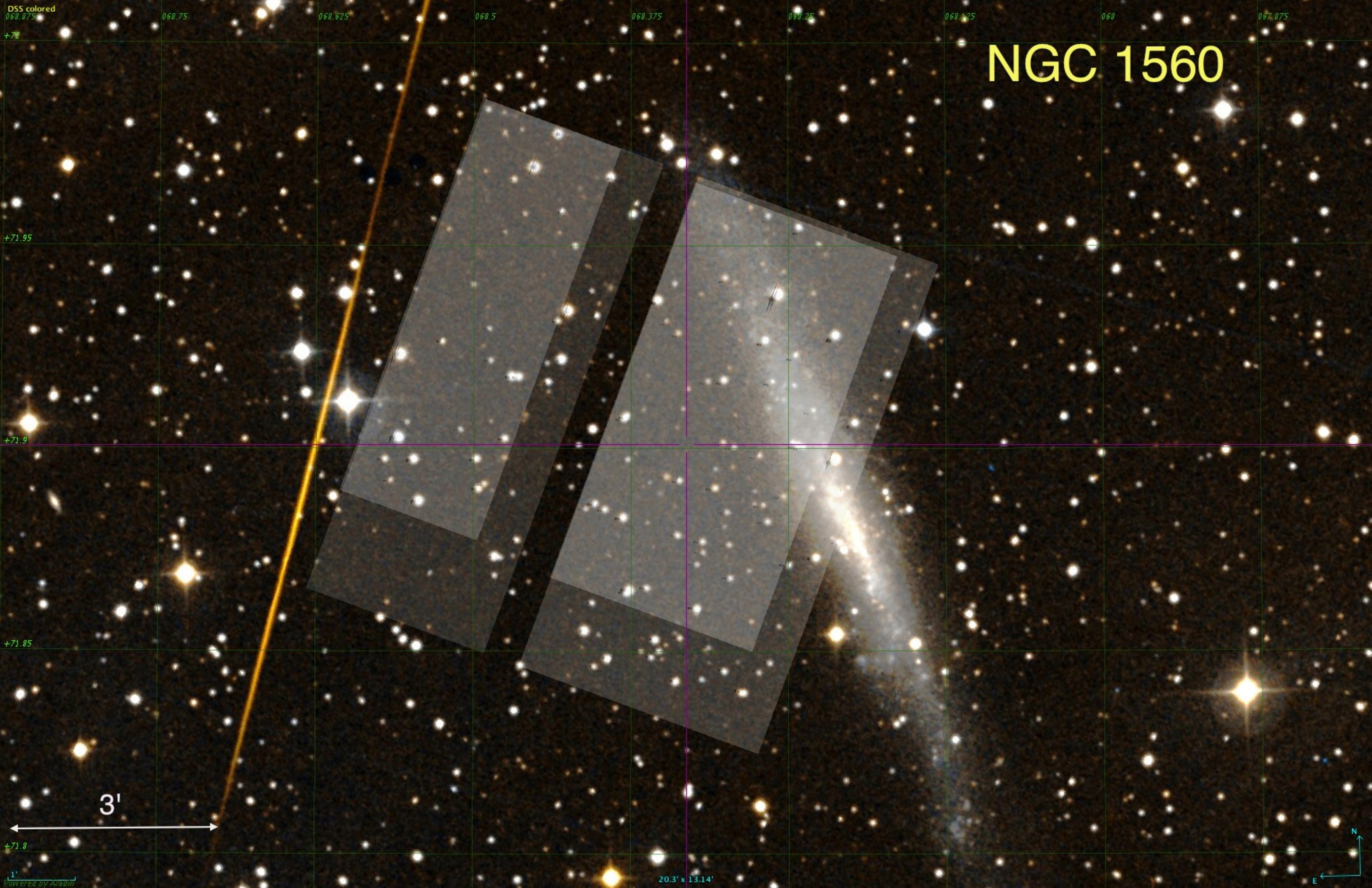}
      \caption{Field observed at GTC superposed onto a color image of the galaxy NGC 1560 from the Digitized Sky Survey. 
      The grey areas represent the final field observed in the $i$ and in the $r$ (light grey) bands . 
      The final usable field of view results from the combination of several individual images with a dithering pattern. In the case of the $r$ band the field is smaller  (see text). The field of view of the DSS image is $\sim$ 19.5 x 13 arcmin$^2$. North is up and East to the left. The scale of 3 arcmin corresponds to a linear scale of 2.9 kpc at the distance of NGC 1560. 
               }
         \label{n1560_fields}
   \end{figure*}

We imaged one field centered at a distance of 3.7 arcmin from the center of the dwarf spiral galaxy NGC 1560 in order to investigate its halo stars (see Figure \ref{n1560_fields}). 
The observations were obtained using the imager OSIRIS 
\footnote{http://www.gtc.iac.es/instruments/osiris/osiris.php}
\citep{cepa2003}   at the  Gran Telescopio CANARIAS (GTC) in La Palma. The imager includes two CCD detectors separated by a 
small gap allowing a total usable field of  7 $\times$ 8 arcmin$^2$. We secured several  short exposure  images of both fields using the Sloan filters $r$ and $i$ with individual exposure time of 60s. In order to improve the quality of the final images  the pointing was changed with a pattern of $\sim$ 5-10 arcsec. In the case of the observations in filter $r$ two slightly different pointings where adopted resulting into a smaller final field of view with respect to that obtained in filter $i$ when combining all frames (see Fig. \ref{n1560_fields}).
All the images were reduced using IRAF \footnote{IRAF (Image Reduction and Analysis Facility) }, including bias subtraction, flat field correction and cleaning of cosmic rays and other minor defects in the detectors. Using the measured positions of a number of relatively bright, unsaturated and isolated stars in the frames  the images were then realigned and co-added.
The parameters of the  final images are reported in Table \ref{table_obs}.
The quality of the the final images is good resulting in PSF with FWHM of 0.85 arcsec in both filters.
Finally we used observations of photometric standard stars to perform the absolute photometric calibration of the frames. 

\begin{figure*}
   \centering
  \includegraphics[width=5.3cm]{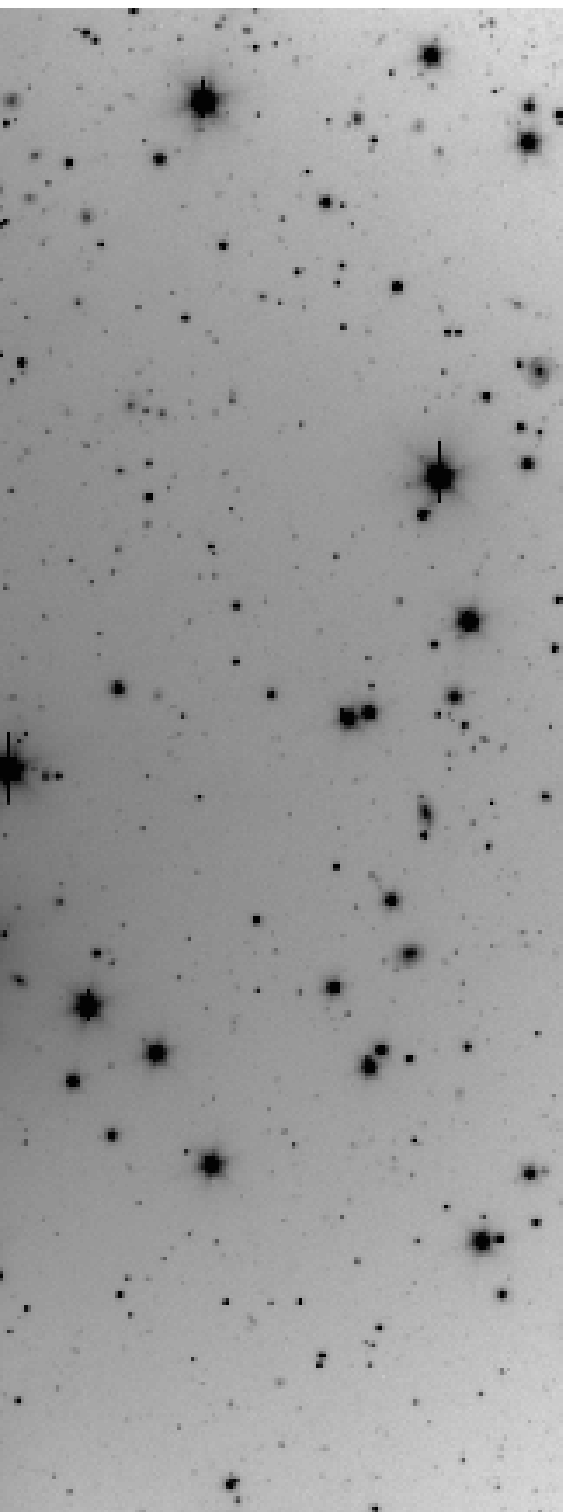}
  \hspace{2cm}
    \includegraphics[width=7.2cm]{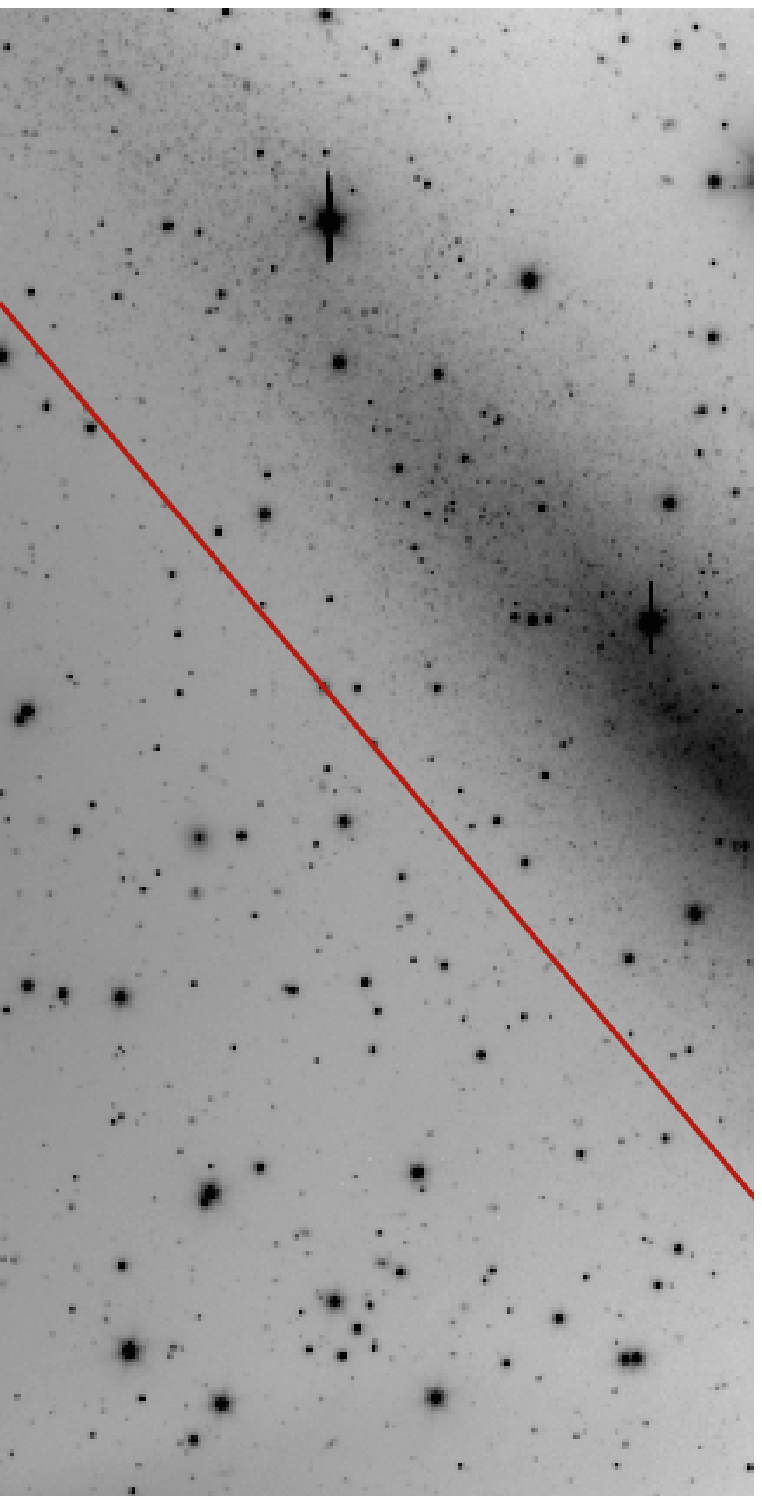}
      \caption{Image in filter $i$ of chip 1 (left) and chip 2 (right panel) observed at GTC close to the galaxy NGC 1560. Filter $i$. Total exposure time is of 1.6 hours. The red line on chip 2 limits the region (lower part of the image) considered for the analysis, in the text referred to as region chip 2a.
               }
         \label{n1560_I1}
   \end{figure*}

\begin{table}
\caption{Properties of the observed fields of NGC 1560}             
\label{table_obs}      
\centering                          
\begin{tabular}{ccccc}        
\hline
Field  & Filter &  FoV & N.of frames & Tot.Exptime \\
         &    &  (arcmin) &  & (sec) \\
\hline
1     &    $i$   &    2.8x7.7    &  90  &     5850  \\
2     &    $i$    &    3.8x7.7    &  90  &     5850  \\
\hline 
1     &    $r$  &    2.1x6.2    &  88  &     5720  \\
2     &    $r$  &    3.2x6.2    &  88  &     5720  \\
\hline
\end{tabular}
\end{table}
The final observed field is shown in Figure \ref{n1560_fields} superimposed on a SDSS image of NGC 1560, while Figure \ref{n1560_I1} shows the coadded $i$ band images of the two chips of the OSIRIS detector.

              \label{fig_fields}

\begin{figure}
\centering
\resizebox{\hsize}{!}{
\includegraphics[angle=0,clip=true]{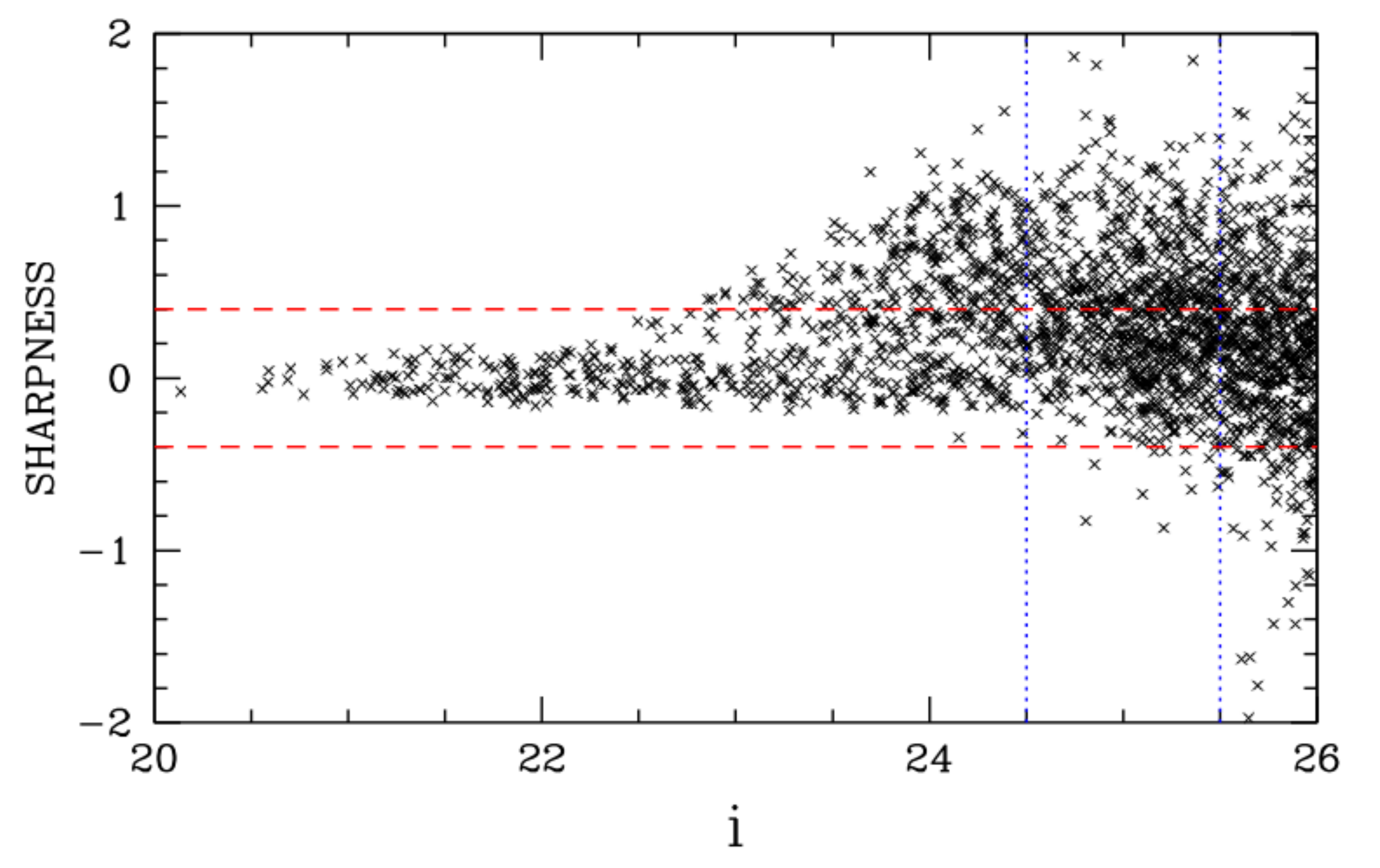}}
   \caption{Sharpness parameter as a function of magnitude for the sources detected in the $i$ band on chip 1 and in chip 2a. The red dashed lines show our criteria adopted to select bona fide stars. The dotted vertical lines show the magnitude range used to map the stars members of NGC 1560, i.e. the upper magnitude on the RGB.}
              \label{fig:sharp_i}
\end{figure}

\begin{figure}
\centering
\resizebox{\hsize}{!}{
\includegraphics[angle=0,clip=true]{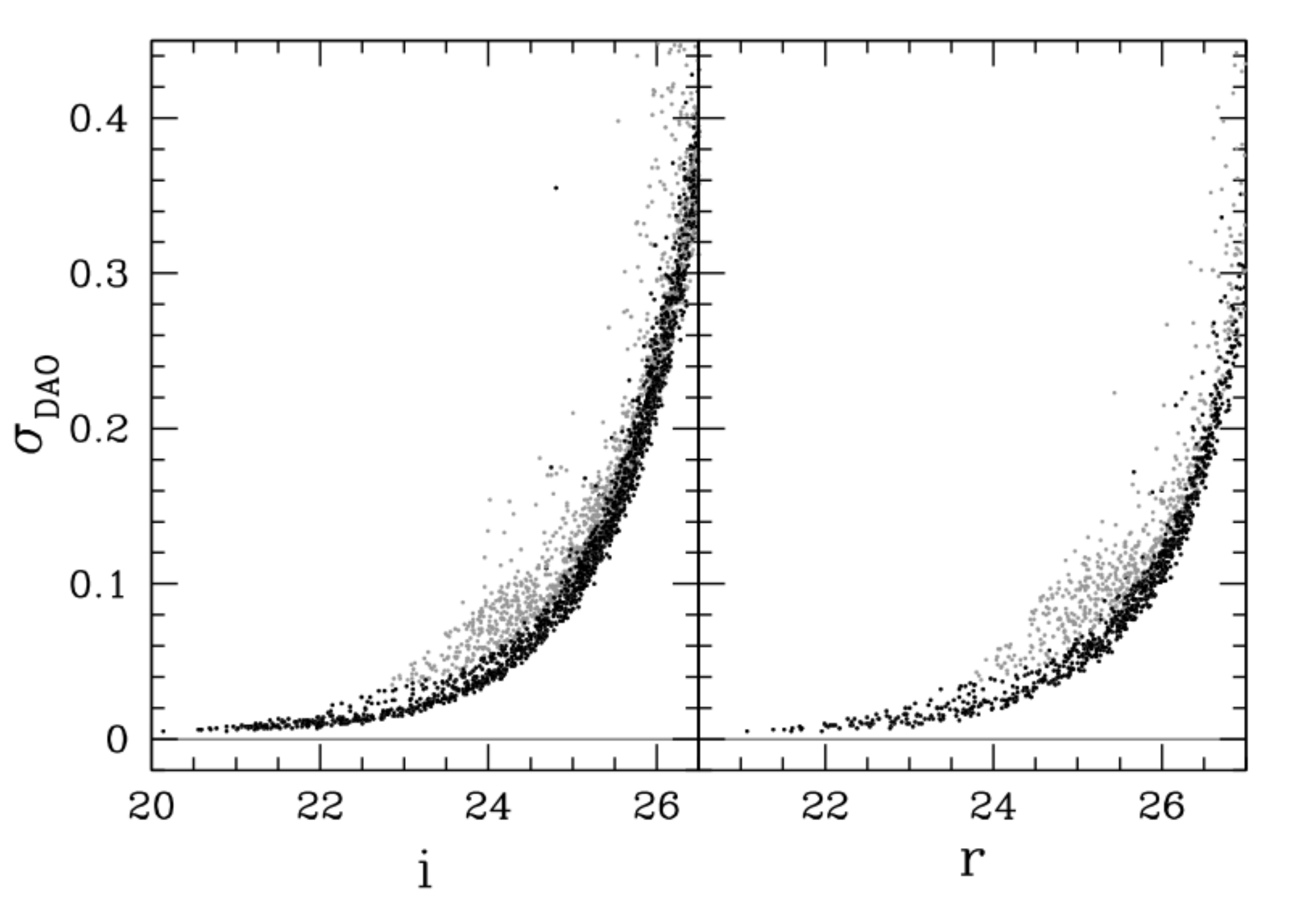}}
   \caption{Photometric error parameter from DAOPHOT ($\sigma_{DAO}$) as a function of magnitude in the $i$ (left) and $r$ (right) bands. The plots show the combination of detections on chip 1 and on chip 2a. Black points highlight sources with sharpness between -0.4 and 0.4. }
             \label{fig:sigmas}
\end{figure}

\begin{figure}
\centering
\resizebox{\hsize}{!}{
\includegraphics[angle=0,clip=true]{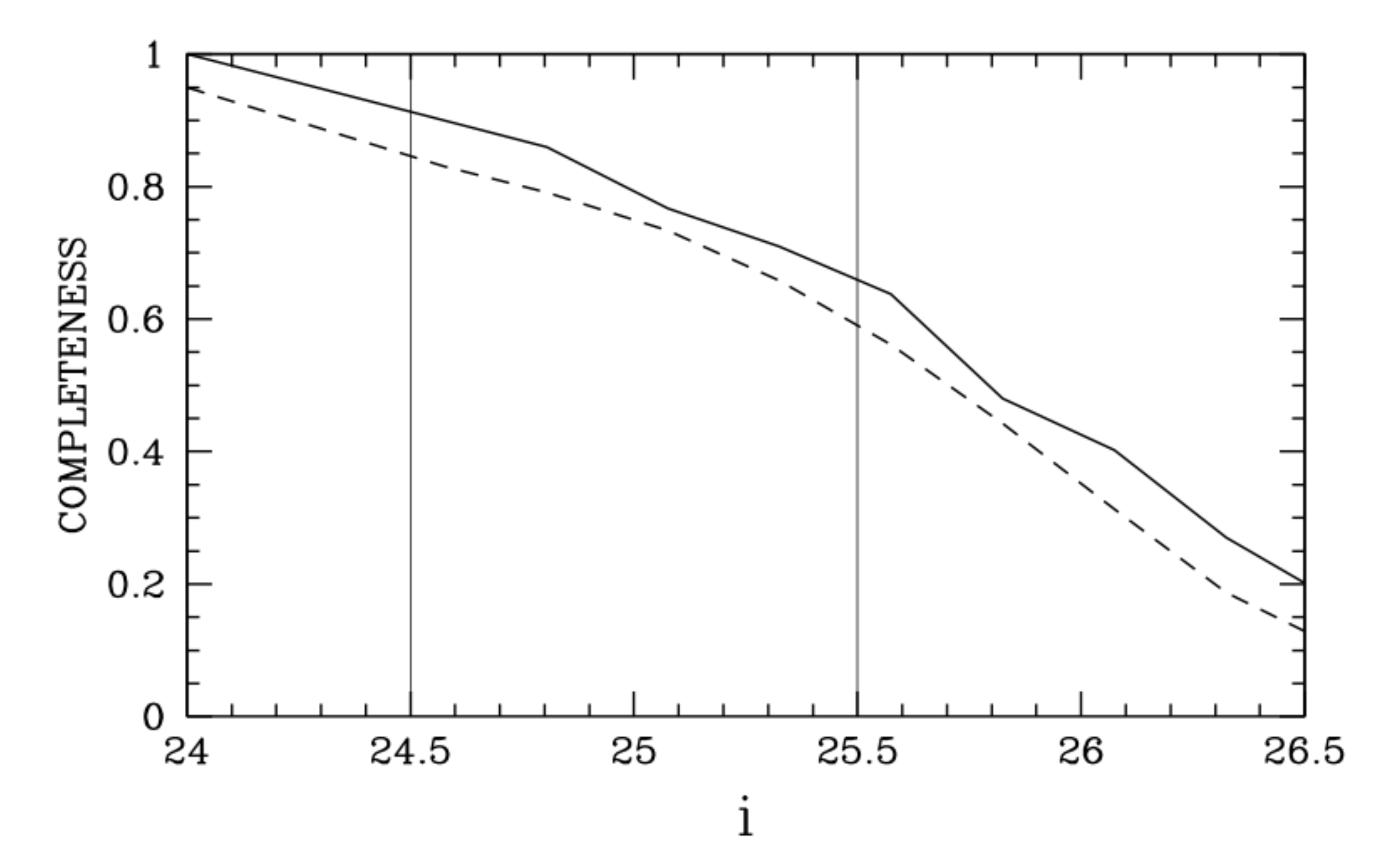}}
   \caption{Fraction of artificial stars recovered and measured with sharpness between -0.4 and +0.4 as function of input magnitude
   on chip 1 (solid) and chip 2a (dashed). The vertical lines show the magnitude of the RGB tip and one magnitude fainter than this. These levels limit the region used to estimate the stellar mass from the star counts.}
               \label{fig:complete}
\end{figure}

\begin{figure*}
\centering
\resizebox{\hsize}{!}{
\includegraphics[angle=0,clip=true]{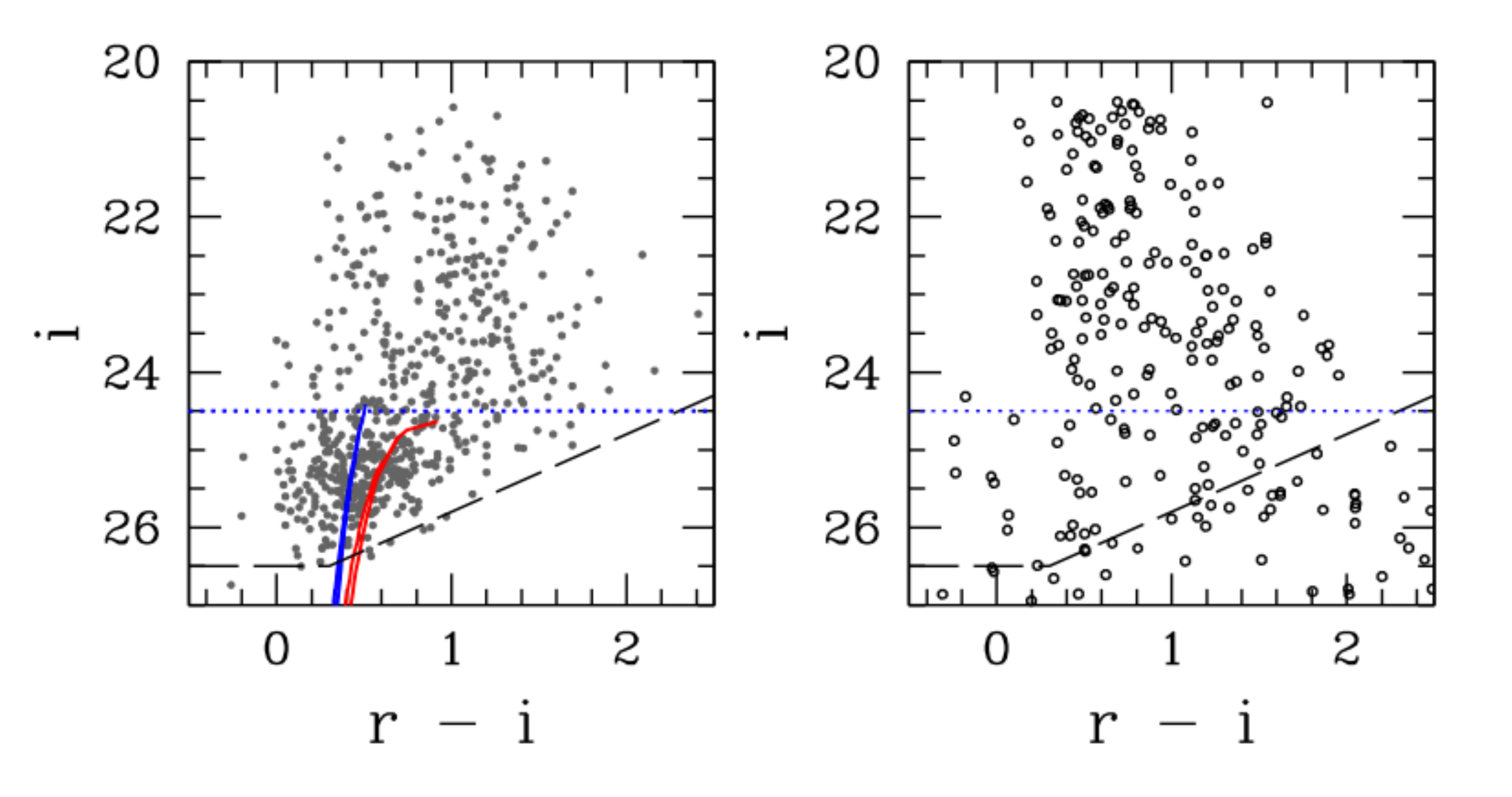}}
   \caption{Left: CMD of the bona fide stars measured on chip 1 and on 
   chip 2a (grey points). The solid lines show 10 Gyr old Padova isochrones for metallicity $Z$ = 0.0004 (blue) and $Z$ = 0.004 (red) at the distance of NGC 1560, including the foreground galactic reddening. Right: CMD of a simulation of the foreground stars in the direction of NGC 1560 for an area close to the surveyed area (see text). Objects brighter than $i$=20.5 are not plotted on this CMD as they do not appear on the observed CMD due to saturation. In both panels the dotted line shows the level of the RGB Tip, and the dashed line shows the magnitude limits of our observations ($i = 26.5$, $r=26.8$). }
               \label{fig:cmd}
\end{figure*}

\begin{figure}
\centering
\resizebox{\hsize}{!}{
\includegraphics[angle=0,clip=true]{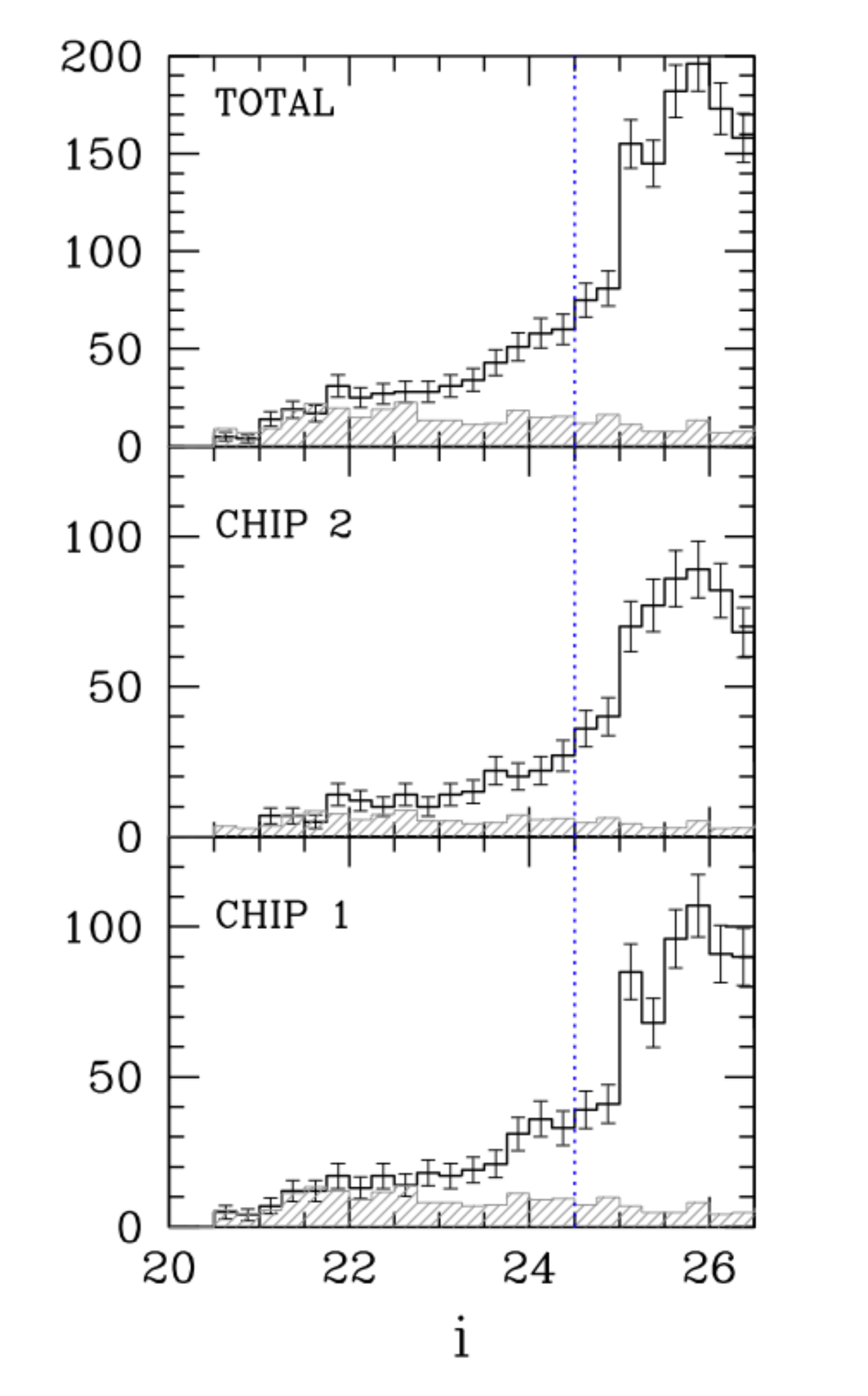}}
   \caption{Magnitude distribution of the bona fide stars on the $i$-band images (unshaded histogram) compared to the TRILEGAL simulation of the foreground stars (shaded histogram). The top panel refers to the total area sampled, while the central and bottom panel refer to the individual chips. The simulations are normalized to the sampled area, respectively of 21.6 and 13.7 square arcmin for chip1 and chip 2a. The vertical dotted line shows the magnitude of the RGB Tip.}
                  \label{fig:lf}
\end{figure}

\begin{figure}
\centering
\resizebox{\hsize}{!}{
\includegraphics[angle=0,clip=true]{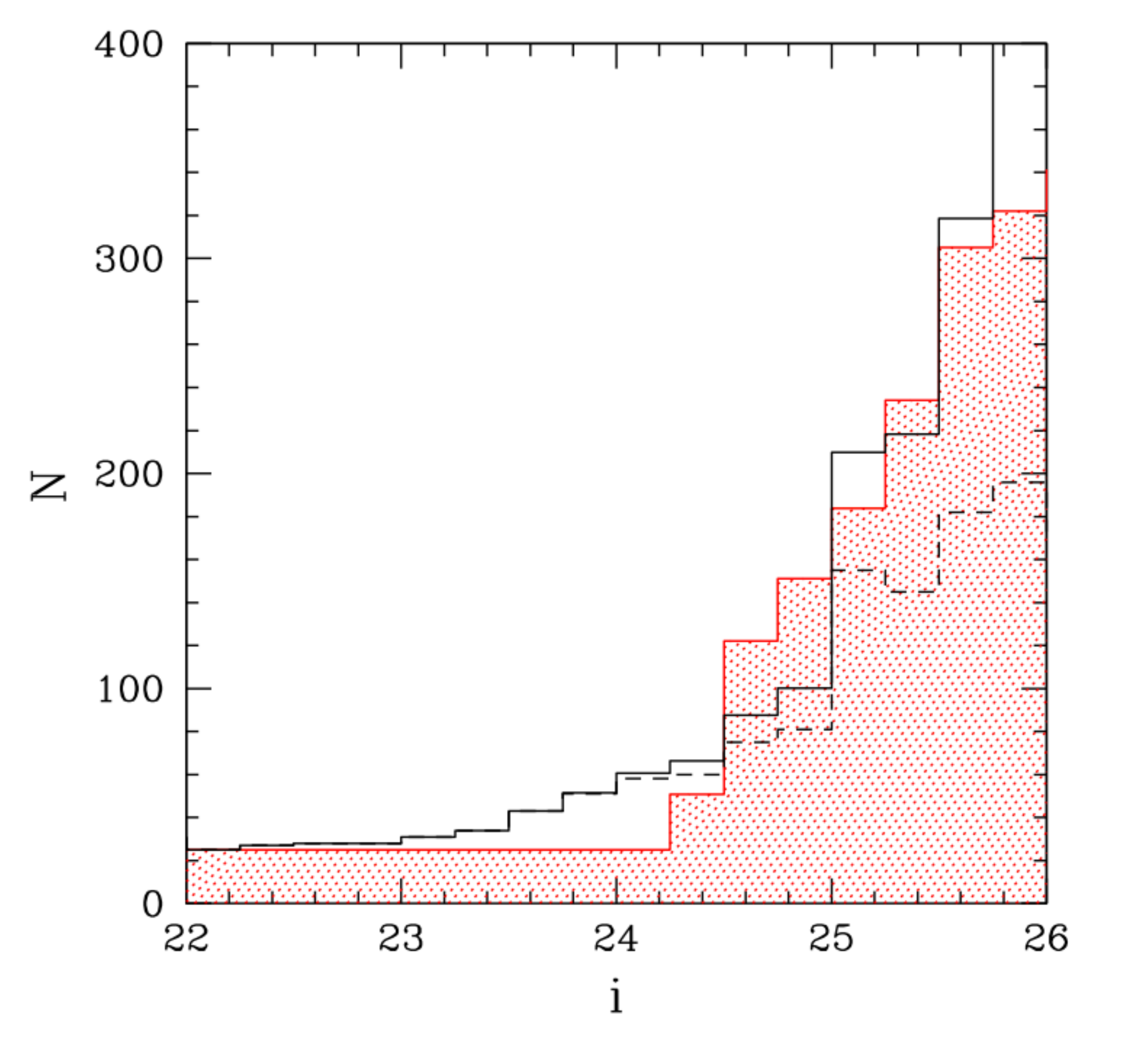}}
   \caption{Magnitude distribution of the bona fide stars on the $i$-band images (unshaded black) compared to the simulation of an
SSP  with an age of 10 Gyr and a metallicity Z=0.0004, based on the Padova tracks (shaded red). The model assumes $(m-M)_0 = 27.57$ and $A_i$ = 0.32.  A foreground stellar density matching the counts in $22 < i < 23$ is included in the model. The observed distribution is plotted as plain counts (dashed lines) and counts corrected for the incompleteness (solid line). Notice that the expected discontinuity at the RGB tip is confused with the component of either foreground or AGB stars at $i \sim$ 24 discussed in  the text.}
                  \label{fig:lf_sim}
\end{figure}

\subsection{Stellar Photometry} \label{sec:photometry}

We performed PSF photometry on the co-added images using the DAOPHOT suite of programs. PSF photometry allowed us to characterize the shape of the sources, so that we could separate stellar from extended objects. Magnitudes were measured with the ALLSTAR routine fitting a PSF constructed with the DAOPSF task, using a number of bright and isolated sources on the individual chips. With a detection threshold of 2.5 $\sigma$, on the $i$ band images we measured 2180 sources on chip 1 and 8837 on chip 2, down to $i \simeq$ 27. In the $r$ band images we measured  1195 and 5435 sources down to $r \simeq$ 27.5 on chip 1 and chip 2, respectively. Chip 2 includes a portion of the disk of NGC 1560, where crowding is high. Since we concentrate on the halo population, in our analysis we only consider sources detected at a distance greater than 1.3  arcmin (corresponding to $\sim$ 1.2 kpc) from the plane of the disk (see Fig. \ref{n1560_I1}), where we count 1619 and 565 sources in the $i$ and $r$ frames respectively.  
In the following, this portion of chip 2 will be referred to as chip 2a. 

We complement our photometric analysis with artificial stars experiments, in which we add  a number of artificial stars of magnitude in the range 24.5 to 26.7 to the $i$-band image. We then
perform photometry of the frame including the artificial stars adopting the same DAOPHOT parameters as done on the original image. These experiments allowed us to refine the criteria to select stellar objects from the quality parameters of the DAOPHOT package ($\chi^2$ and sharpness) , as well as to assess the error (
as difference between the input and output magnitude) and the completeness of detected stars as a function of their magnitude.

Based on the distribution of the $\chi^2$ parameter we discarded the objects with $\chi^2 \geq$ 0.5. This criterion excludes $\sim$ 10 
\% of the sources; most of them are bright, saturated foreground stars. Fig. \ref{fig:sharp_i} shows the distribution of the sharpness 
parameter as a function of the magnitude of the sources, for objects brighter than $i=26$, that is $\sim$ 1.5 mag fainter than the expected location of the RGB Tip.
At $i \lesssim 24$ most of the sources have sharpness between $-$0.2 and 0.2, but  at fainter magnitudes the spread grows. At $i \gtrsim 24$ a large number of sources with high and positive sharpness values appear.
Guided by the results of the artificial stars experiments we selected as bona fide stars the sources with sharpness smaller than 0.4. Upon visual inspection, most sources with sharpness higher than this appear as extended objects. The lower limit to the sharpness parameter adopted is $-$0.4. The total number of bona fide stars are 983 and 789 respectively on chip 1 and on chip 2a. The distributions of the $\chi^2$ and sharpness parameters on the $r$-band images have very similar characteristics as those of the $i$-band images. The bona fide stars are 539 in chip 1 and 340 in chip 2a.

Fig. \ref{fig:sigmas} shows the DAOPHOT photometric accuracy parameter ($\sigma_{DAO}$) as function of magnitude for the $i$ and $r$ bands in the two chips, having again excluded the disk of NGC 1560.  The adopted criteria to select the bona fide stars also select the best measured objects.  

We use the results of the artificial stars experiments to evaluate the completeness of our data as function of magnitude by comparing the number of stars measured on the images with the synthetic stars added,  to the number of input artificial stars.
Fig. \ref{fig:complete} shows the completeness factors as a function of the magnitude. We note that in the  magnitude range of interest the completeness of our data is better than 50 \% for the bona fide selected sources. The artificial stars experiments were also used to evaluate the true error on the measured magnitude of the bona fide stars. Down to $i \sim 26$ the distribution of $\Delta {\rm m} = {\rm m}_{output} - {\rm m}_{input}$ is found symmetric around 0 with a dispersion of 0.1, 0.15 and 0.2 for $i \simeq 25, 25.5$ and 26 respectively. At fainter magnitudes the distribution becomes noticeably skewed towards negative values of   $\Delta {\rm m}$ because sources artificially brightened by blending are more likely to be detected than sources whose brightness is underestimated \citep[see][]{Book11}. We conclude that in the range $i \lesssim 26$ the quality of our photometry is satisfactory.

\section{The Stellar Halo around NGC 1560} \label{sec:CMD}

Matching the sources detected in both $r$ and $i$ we get 302 and 284 bona fide stars respectively on chip 1 and on  chip 2a.  
Fig. \ref{fig:cmd} shows the relative CMD, to which we superpose 10 Gyr old Padova isochrones with two subsolar metallicities. We compare this CMD with the result of a simulation of the foreground contamination obtained using the TRILEGAL\footnote{http://stev.oapd.inaf.it/cgi-bin/trilegal} tool \citep{Girardi+05}, at the galactic coordinates of NGC 1560 (see Fig. \ref{fig:cmd} , right panel). The simulation is computed  for an area of 25 arcmin$^2$, close to the area on which our observed CMD is constructed.  At magnitudes brighter than our fiducial RGB Tip the observed CMD is very similar to the simulated one, but fainter than RGB Tip there is a clear excess of observed stars with colors appropriate to those of a metal poor old stellar population. The expected number of foreground stars at  $i  > 24.5$ is significantly smaller than that observed. In addition, at these magnitudes the simulated stars  
scatter over a much wider color range compared to the data. In the range $24.5 < i < 26$ and $ 0 < r - i < 1$ we count 274 stars on our CMD, while the simulation of the foreground component contains only  17 objects.  We also computed a simulation of the foreground stars using the recently updated Besancon model \citep{Robin+03}\footnote{http://model2016.obs-besancon.fr/} which yields 12 expected foreground objects in this region of the CMD , i.e. even less than the TRILEGAL simulation.This indicates that the observed CMD contains a population of stars members of NGC 1560. In the following, for the foreground stars contamination we will use the more conservative value obtained from TRILEGAL. 

The star counts on the $i$ band image shown in Fig. \ref{fig:lf} further support our detection of the NGC 1560 stellar halo . 
Compared to the foreground, the observed star counts are close to the expected level at  $i < $ 23, while they show a mild excess (a factor of $\gtrsim$ 2) at $i \sim 24$. These excess sources could be either bright Asymptotic Giant Branch (AGB) stars members of the NGC 1560 halo, or  Milky Way stars, if the foreground contribution at these magnitudes were underestimated in the adopted Galaxy model.  At $25 < i < 26$ a remarkable (a factor of $\sim$ 10) difference between the observed counts and the foreground level is apparent (see Fig. \ref{fig:lf}). This large excess is present in both chips: the surface density of bona fide stars with  $24.5 \leq i \leq 25.5$ is 11  per square arcmin in chip 1, 16 per square arcmin in chip 2a, to be compared with the 1.5 objects per square arcmin of the TRILEGAL simulation. Therefore, most of the stars fainter than $i \simeq 24.5$  belong to the stellar halo of NGC 1560. 

In order to estimate the stellar mass traced by our halo field we construct a simple stellar population (SSP) model by distributing stars along an old and metal poor isochrone, following a prescribed initial mass function (IMF). We adopt a 10 Gyr old isochrone with a metallicity of $Z = Z_{\odot}/50$ from the Padova \citep{Marigo+08} set. The choice of  the metallicity is motivated by the color of the RGB stars in our field (see Fig. \ref{fig:cmd}), while the adopted value of the age is not critical, since the conversion factor between bright RGB star counts and mass of the parent stellar population is almost constant for ages older than  $\sim$ 2 Gyr \citep{Book11}. 
Figure \ref{fig:lf_sim} shows the magnitude distribution of this synthetic stellar population, compared to 
our data. The model is in acceptable agreement with  the completeness corrected observed counts.
The normalization of the  SSP model yields the total stellar mass needed to reproduce the observed star counts, which turns out of  $\sim 1.6 \times 10^7$ \msun, having adopted a Salpeter IMF ( between 0.1 and 100 \msun). This is the total mass transformed into stars, while the current mass will be a factor $\sim$ 0.7 lower because of the mass returned to the interstellar medium by  stellar wind and supernovae over the long lifetime of the stellar population \citep[see][]{Book11}.  

Assuming a spherical halo with a surface mass density described by a power law of index of $s = -1.5$, close to what found for NGC 253 in \cite{Greggio+14}, we find that our $i$ band image samples $\sim$ 13 \% of the total mass.  This indicates  a value of  $ \sim 10^8$ \msun\ stellar halo for this galaxy. Varying the power law index, the mass fraction sampled by our field is of  13 \% for 
$s \lesssim -1.5$,  of 10\% for $s=-2$, and decreases to  $\sim$ 4 \% as the profile steepens to $s = -3$. 
Since chip 1 samples the halo in an outer region compared to chip 2, we may derive indications on the slope of the mass distribution by comparing the star counts in the two chips. The number of bona fide stars with $24.5\leq i \leq 25.5$ is 234 in chip 1 and 223 in chip 2a . Upon correction  for incompleteness, these figures become respectively 332 and 348. The counts include the halo members of NGC 1560 as well as the foreground stars, which, according to the TRILEGAL model, should amount to $\sim$ 40 stars in chip 1 and   25 stars on chip 2 in the same magnitude range. The true members on NGC 1560 should therefore be 292 and 323 in chip 1 and 2a respectively. On the other hand, as argued above, the TRILEGAL model matches very well the star counts in the range $22 \lesssim i \lesssim 23$, but it falls short of the observed counts in $23 \lesssim i \lesssim 24.5$, where we do not expect a sizable population of  NGC 1560 members. In fact, these stars should be bright AGB objects, which are very rare in stellar populations older than a few Gyrs \citep[e.g.][]{Noel+13}.
If we estimate the foreground contribution from the counts at magnitudes just brighter than the RGB tip, ( $i \simeq 24.3$), chip 1 and chip 2a should include respectively $\sim 120$ and 76 foreground stars with $24.5 < i < 25.5$. 
 This would leave 212 and 272 members of NGC 1560 halo stars sampled in chip 1 and 2a. From this discussion we estimate that the ratio between the NGC 1560 halo stars in chip 1 and in chip 2a varies between 0.9 and 0.78, depending on the level of the foreground contribution to the counts. 
For a spherical halo with a surface mass density scaling as a power law the ratio of the mass sampled in chip 1 and chip 2a decreases from 0.7 to 0.13 as the index varies from $-1$ to $-3$.  Therefore our data  suggest a quite shallow profile of the mass distribution for the halo of NGC 1560.

\section{Summary and discussion} \label{sec:summary}

We performed deep photometry  in a region centered at a projected distance of $\sim$ 3.5 kpc  from the center  NGC 1560, a low mass, late type spiral galaxy at a distance of 3.27 Mpc. The favorable edge on orientation of this galaxy, coupled with the relatively wide FoV of the observations allowed us to derive a fair census of the sources projected on the halo area around NGC 1560, where we detect a large excess of stars above the expected foreground contribution. Based on the colors and magnitudes of these sources we conclude that they are bright RGB stars nembers of NGC 1560. Their $(r-i)$ color is appropriate for a metal poor stellar population ([Fe/H] $\simeq -1.7$). We measure this excess sources  all over the surveyed region, with no apparent concentration towards the galaxy disk.
Therefore we conclude that \textit{these stars belong to an extended stellar halo around NGC 1560}.

The size of the excess counts indicates a stellar mass of $\sim 10^7$ \msun\ in the sampled region, assuming a Salpeter IMF. Comparing the counts in two regions of our surveyed  area we favor a quite shallow mass surface density profile of the halo component, with a slope of $\sim -1$ 
for a power law distribution. Assuming a spherical halo with such a shallow profile, and keeping in mind the uncertainty of the foreground contribution, the sampled region collects 13 \% of the total halo mass, which then turns out of $\sim$ $10^{8} \msun$. 

The sampled mass and the profile that we derive depend on the assumptions regarding the contribution of the foreground stars and of the background galaxies. Both the TRILEGAL and the Besancon simulators of the Milky Way population in the direction of NGC 1560  predict a negligible number of foreground sources at magnitudes fainter than the RGB Tip in our sampled fields. On the other hand, the number of compact background galaxies at these faint magnitudes might be relevant. To minimize this kind of contaminants we have selected sources with a small value of the sharpness parameter from the DAOPHOT PSF fitting photometry.  Based on the results of the artificial stars experiments, we defined as bona fide stars sources with sharpness smaller than  0.4 (in absolute value). 
Using a more restrictive criterion the mass sampled on chip 1 and chip 2a varies. For example, changing the threshold from 0.4 to 0.2, the number of bona fide stars decreases by $\sim$ 25 \% in both chips. While this corresponds a smaller mass for the stellar halo of NGC 1560, the excess of point like sources with respect to the expected foreground stars remains significant. The detection of an extended stellar halo around NGC 1560 is thus a robust result, while the measurement of its mass and profile is subject to some uncertainty depending on the contribution of unresolved compact galaxies.
 
 We estimate the stellar mass of the main body of NGC 1560 from its I band magnitude ($I = 10.26$, \cite{Buta+99}), adopting a Milky Way absorption  of  $A_{I} = 0.28$, and a mass-to-light ratio of $M/L_{I} = 1.7$ in solar units, which is appropriate for a stellar population with a constant star formation rate over the Hubble time, and a Salpeter IMF. With this assumptions, and our adopted distance modulus of 27.6, the mass NGC 1560 results of $8 \times 10^{8} \msun$, so that the ratio between the mass of the halo and the mass of the main body of the galaxy is $\sim$ 0.1.

\cite{Harmsen+17} summarize the properties of  the stellar haloes of Milky Way mass spiral galaxies from the GHOSTS survey. These appear characterized by steep surface mass density profiles ($s \sim -2 \div -3.7$), and a large diversity of stellar halo mass fractions ($\sim 0.01 \div 0.1$). Our data for NGC 1560 suggest a quite shallower profile, and a high stellar halo mass fraction. How do these properties compare to the results of numerical simulations?

Detailed computations of the structure of galaxy stellar haloes within the $\Lambda$ cold dark matter cosmological model are available mostly for galaxies with mass in a range akin to the Milky Way and Andromeda \citep[e.g.][]{ Lackner+12, McCarthy+12, Cooper+13, Tissera+14}. Recently \cite{Rodriguez+16} have extended the mass range to include objects with stellar mass down to $10^9$ \msun.  In their models, the lowest mass galaxies typically host an accreted stellar mass of few percent of the total galaxy mass, but with a large variance, and an accreted mass fraction up to 0.1 is reported for galaxies of $\sim 10^9 \msun$. In these models,   
the accreted stars are distributed in an elongated envelope characterized by a steep density drop. 
Steep slopes for the outer regions are also found in models of dwarf galaxies \citep{Stinson+09, Bekki08, Valcke+08}, and  in observational data \citep[e.g.][]{Battinelli+06, Dsouza+14}. 
Although a direct comparison of our results with those of other galaxies, and of theoretical models, is hindered by differences in mass and methods of analysis, it appears that in the case of NGC 1560, the stellar halo has an extended and flat profile  
which may pose important constraints to galaxy formation models of low mass spiral galaxies. This halo extends over a wide region: we detect stars at distances 
3 $\times$ larger than the edge of the HI disk, as measured by \citet{Gentile+10}. The HI distribution appears undisturbed, the galaxy is isolated, and there is no evidence of a recent interaction.
This suggests that the stellar halo of NGC 1560 was acquired at an early epoch, possibly built during the original collapse which led to the formation of the disk. 

\acknowledgments

The financial contribution by the contract \textit{Studio e Simulazioni di Osservazioni (Immagini e Spettri) con MICADO per E-ELT} (DD 27/2016 - Ob. Fun. 1.05.02.17)  of the INAF project \textit{Micado simulazioni casi scientifici}, P.I. Dr. Renato Falomo,  is acknowledged.

This research has made use of the NASA/IPAC Extragalactic Database (NED) which is operated by the Jet Propulsion Laboratory, California Institute of Technology, under contract with the National Aeronautics and Space Administration.

\facilities{GTC-OSIRIS, \citep{cepa2003}}

\bibliographystyle{aasjournal}
\bibliography{N1560_biblio}

\begin{thebibliography}{}
\expandafter\ifx\csname natexlab\endcsname\relax\def\natexlab#1{#1}\fi

\bibitem[{{Barker} {et~al.}(2009){Barker}, {Ferguson}, {Irwin}, {Arimoto}, \&
  {Jablonka}}]{Barker+09}
{Barker}, M.~K., {Ferguson}, A.~M.~N., {Irwin}, M., {Arimoto}, N., \&
  {Jablonka}, P. 2009, \aj, 138, 1469

\bibitem[{{Battinelli} {et~al.}(2006){Battinelli}, {Demers}, \&
  {Kunkel}}]{Battinelli+06}
{Battinelli}, P., {Demers}, S., \& {Kunkel}, W.~E. 2006, \aap, 451, 99

\bibitem[{{Bekki}(2008)}]{Bekki08}
{Bekki}, K. 2008, \apjl, 680, L29

\bibitem[{{Buta} \& {McCall}(1999)}]{Buta+99}
{Buta}, R.~J., \& {McCall}, M.~L. 1999, \apjs, 124, 33

\bibitem[{{Cepa} {et~al.}(2003){Cepa}, {Aguiar-Gonzalez}, {Bland-Hawthorn},
  {Castaneda}, {Cobos}, {Correa}, {Espejo}, {Fragoso-Lopez}, {Fuentes},
  {Gigante}, {Gonzalez}, {Gonzalez-Escalera}, {Gonzalez-Serrano},
  {Joven-Alvarez}, {Lopez-Ruiz}, {Militello}, {Cano}, {Perez}, {Perez},
  {Rasilla}, {Sanchez}, \& {Tejada}}]{cepa2003}
{Cepa}, J., {Aguiar-Gonzalez}, M., {Bland-Hawthorn}, J., {et~al.} 2003, in
  Proc. SPIE, Vol. 4841, -, 1739--1749

\bibitem[{{Cooper} {et~al.}(2013){Cooper}, {D'Souza}, {Kauffmann}, {Wang},
  {Boylan-Kolchin}, {Guo}, {Frenk}, \& {White}}]{Cooper+13}
{Cooper}, A.~P., {D'Souza}, R., {Kauffmann}, G., {et~al.} 2013, \mnras, 434,
  3348

\bibitem[{{Cooper} {et~al.}(2010){Cooper}, {Cole}, {Frenk}, {White}, {Helly},
  {Benson}, {De Lucia}, {Helmi}, {Jenkins}, {Navarro}, {Springel}, \&
  {Wang}}]{Cooper+10}
{Cooper}, A.~P., {Cole}, S., {Frenk}, C.~S., {et~al.} 2010, \mnras, 406, 744

\bibitem[{{D'Souza} {et~al.}(2014){D'Souza}, {Kauffman}, {Wang}, \&
  {Vegetti}}]{Dsouza+14}
{D'Souza}, R., {Kauffman}, G., {Wang}, J., \& {Vegetti}, S. 2014, \mnras, 443,
  1433

\bibitem[{{Ferguson} {et~al.}(2002){Ferguson}, {Irwin}, {Ibata}, {Lewis}, \&
  {Tanvir}}]{Ferguson+02}
{Ferguson}, A.~M.~N., {Irwin}, M.~J., {Ibata}, R.~A., {Lewis}, G.~F., \&
  {Tanvir}, N.~R. 2002, \aj, 124, 1452

\bibitem[{{Foster} {et~al.}(2014){Foster}, {Lux}, {Romanowsky},
  {Mart{\'{\i}}nez-Delgado}, {Zibetti}, {Arnold}, {Brodie}, {Ciardullo},
  {GaBany}, {Merrifield}, {Singh}, \& {Strader}}]{Foster+14}
{Foster}, C., {Lux}, H., {Romanowsky}, A.~J., {et~al.} 2014, \mnras, 442, 3544

\bibitem[{{Gentile} {et~al.}(2010){Gentile}, {Baes}, {Famaey}, \& {van
  Acoleyen}}]{Gentile+10}
{Gentile}, G., {Baes}, M., {Famaey}, B., \& {van Acoleyen}, K. 2010, \mnras,
  406, 2493

\bibitem[{{Girardi} {et~al.}(2005){Girardi}, {Groenewegen}, {Hatziminaoglou},
  \& {da Costa}}]{Girardi+05}
{Girardi}, L., {Groenewegen}, M.~A.~T., {Hatziminaoglou}, E., \& {da Costa}, L.
  2005, \aap, 436, 895

\bibitem[{{Greggio} {et~al.}(2014){Greggio}, {Rejkuba}, {Gonzalez},
  {Arnaboldi}, {Iodice}, {Irwin}, {Neeser}, \& {Emerson}}]{Greggio+14}
{Greggio}, L., {Rejkuba}, M., {Gonzalez}, O.~A., {et~al.} 2014, \aap, 562, A73

\bibitem[{{Greggio} \& {Renzini}(2011)}]{Book11}
{Greggio}, L., \& {Renzini}, A. 2011, {Stellar Populations. A User Guide from
  Low to High Redshift}

\bibitem[{{Harmsen} {et~al.}(2017){Harmsen}, {Monachesi}, {Bell}, {de Jong},
  {Bailin}, {Radburn-Smith}, \& {Holwerda}}]{Harmsen+17}
{Harmsen}, B., {Monachesi}, A., {Bell}, E.~F., {et~al.} 2017, \mnras, 466, 1491

\bibitem[{{Jacobs} {et~al.}(2009){Jacobs}, {Rizzi}, {Tully}, {Shaya},
  {Makarov}, \& {Makarova}}]{Jacobs+09}
{Jacobs}, B.~A., {Rizzi}, L., {Tully}, R.~B., {et~al.} 2009, \aj, 138, 332

\bibitem[{{Karachentsev} {et~al.}(2004){Karachentsev}, {Karachentseva},
  {Huchtmeier}, \& {Makarov}}]{Karachen+04}
{Karachentsev}, I.~D., {Karachentseva}, V.~E., {Huchtmeier}, W.~K., \&
  {Makarov}, D.~I. 2004, \aj, 127, 2031

\bibitem[{{Lackner} {et~al.}(2012){Lackner}, {Cen}, {Ostriker}, \&
  {Joung}}]{Lackner+12}
{Lackner}, C.~N., {Cen}, R., {Ostriker}, J.~P., \& {Joung}, M.~R. 2012, \mnras,
  425, 641

\bibitem[{{Marigo} {et~al.}(2008){Marigo}, {Girardi}, {Bressan}, {Groenewegen},
  {Silva}, \& {Granato}}]{Marigo+08}
{Marigo}, P., {Girardi}, L., {Bressan}, A., {et~al.} 2008, \aap, 482, 883

\bibitem[{{Mart{\'{\i}}nez-Delgado} {et~al.}(2015){Mart{\'{\i}}nez-Delgado},
  {D'Onghia}, {Chonis}, {Beaton}, {Teuwen}, {GaBany}, {Grebel}, \&
  {Morales}}]{Martinez+15}
{Mart{\'{\i}}nez-Delgado}, D., {D'Onghia}, E., {Chonis}, T.~S., {et~al.} 2015,
  \aj, 150, 116

\bibitem[{{Mart{\'{\i}}nez-Delgado} {et~al.}(2010){Mart{\'{\i}}nez-Delgado},
  {Gabany}, {Crawford}, {Zibetti}, {Majewski}, {Rix}, {Fliri},
  {Carballo-Bello}, {Bardalez-Gagliuffi}, {Pe{\~n}arrubia}, {Chonis}, {Madore},
  {Trujillo}, {Schirmer}, \& {McDavid}}]{Martinez+10}
{Mart{\'{\i}}nez-Delgado}, D., {Gabany}, R.~J., {Crawford}, K., {et~al.} 2010,
  \aj, 140, 962

\bibitem[{{Mart{\'{\i}}nez-Delgado} {et~al.}(2012){Mart{\'{\i}}nez-Delgado},
  {Romanowsky}, {Gabany}, {Annibali}, {Arnold}, {Fliri}, {Zibetti}, {van der
  Marel}, {Rix}, {Chonis}, {Carballo-Bello}, {Aloisi}, {Macci{\`o}},
  {Gallego-Laborda}, {Brodie}, \& {Merrifield}}]{Martinez+12}
{Mart{\'{\i}}nez-Delgado}, D., {Romanowsky}, A.~J., {Gabany}, R.~J., {et~al.}
  2012, \apjl, 748, L24

\bibitem[{{McCarthy} {et~al.}(2012){McCarthy}, {Font}, {Crain}, {Deason},
  {Schaye}, \& {Theuns}}]{McCarthy+12}
{McCarthy}, I.~G., {Font}, A.~S., {Crain}, R.~A., {et~al.} 2012, \mnras, 420,
  2245

\bibitem[{{McConnachie}(2016)}]{McConnachie16}
{McConnachie}, A.~W. 2016, in IAU Symposium, Vol. 317, The General Assembly of
  Galaxy Halos: Structure, Origin and Evolution, ed. A.~{Bragaglia},
  M.~{Arnaboldi}, M.~{Rejkuba}, \& D.~{Romano}, 15--20

\bibitem[{{Merritt} {et~al.}(2016){Merritt}, {van Dokkum}, {Abraham}, \&
  {Zhang}}]{Merritt+16}
{Merritt}, A., {van Dokkum}, P., {Abraham}, R., \& {Zhang}, J. 2016, \apj, 830,
  62

\bibitem[{{Minniti} \& {Zijlstra}(1996)}]{Minniti+96}
{Minniti}, D., \& {Zijlstra}, A.~A. 1996, \apjl, 467, L13

\bibitem[{{Miskolczi} {et~al.}(2011){Miskolczi}, {Bomans}, \&
  {Dettmar}}]{Miskolczi+11}
{Miskolczi}, A., {Bomans}, D.~J., \& {Dettmar}, R.-J. 2011, \aap, 536, A66

\bibitem[{{Mouhcine} {et~al.}(2010){Mouhcine}, {Ibata}, \&
  {Rejkuba}}]{Mouhcine+10}
{Mouhcine}, M., {Ibata}, R., \& {Rejkuba}, M. 2010, \apjl, 714, L12

\bibitem[{{Mouhcine} {et~al.}(2005){Mouhcine}, {Rich}, {Ferguson}, {Brown}, \&
  {Smith}}]{Mouhcine+05}
{Mouhcine}, M., {Rich}, R.~M., {Ferguson}, H.~C., {Brown}, T.~M., \& {Smith},
  T.~E. 2005, \apj, 633, 828

\bibitem[{{No{\"e}l} {et~al.}(2013){No{\"e}l}, {Greggio}, {Renzini}, {Carollo},
  \& {Maraston}}]{Noel+13}
{No{\"e}l}, N.~E.~D., {Greggio}, L., {Renzini}, A., {Carollo}, C.~M., \&
  {Maraston}, C. 2013, \apj, 772, 58

\bibitem[{{Rich} {et~al.}(2012){Rich}, {Collins}, {Black}, {Longstaff}, {Koch},
  {Benson}, \& {Reitzel}}]{Rich+12}
{Rich}, R.~M., {Collins}, M.~L.~M., {Black}, C.~M., {et~al.} 2012, \nat, 482,
  192

\bibitem[{{Robin} {et~al.}(2003){Robin}, {Reyl{\'e}}, {Derri{\`e}re}, \&
  {Picaud}}]{Robin+03}
{Robin}, A.~C., {Reyl{\'e}}, C., {Derri{\`e}re}, S., \& {Picaud}, S. 2003,
  \aap, 409, 523

\bibitem[{{Rodriguez-Gomez} {et~al.}(2016){Rodriguez-Gomez}, {Pillepich},
  {Sales}, {Genel}, {Vogelsberger}, {Zhu}, {Wellons}, {Nelson}, {Torrey},
  {Springel}, {Ma}, \& {Hernquist}}]{Rodriguez+16}
{Rodriguez-Gomez}, V., {Pillepich}, A., {Sales}, L.~V., {et~al.} 2016, \mnras,
  458, 2371

\bibitem[{{Stinson} {et~al.}(2009){Stinson}, {Dalcanton}, {Quinn}, {Gogarten},
  {Kaufmann}, \& {Wadsley}}]{Stinson+09}
{Stinson}, G.~S., {Dalcanton}, J.~J., {Quinn}, T., {et~al.} 2009, \mnras, 395,
  1455

\bibitem[{{Tanaka} {et~al.}(2011){Tanaka}, {Chiba}, {Komiyama}, {Guhathakurta},
  \& {Kalirai}}]{Tanaka+11}
{Tanaka}, M., {Chiba}, M., {Komiyama}, Y., {Guhathakurta}, P., \& {Kalirai},
  J.~S. 2011, \apj, 738, 150

\bibitem[{{Tissera} {et~al.}(2014){Tissera}, {Beers}, {Carollo}, \&
  {Scannapieco}}]{Tissera+14}
{Tissera}, P.~B., {Beers}, T.~C., {Carollo}, D., \& {Scannapieco}, C. 2014,
  \mnras, 439, 3128

\bibitem[{{Valcke} {et~al.}(2008){Valcke}, {de Rijcke}, \&
  {Dejonghe}}]{Valcke+08}
{Valcke}, S., {de Rijcke}, S., \& {Dejonghe}, H. 2008, \mnras, 389, 1111

\end{thebibliography}

\end{document}